\documentclass[aps,pre,reprint, floatfix,showpacs,superscriptaddress,longbibliography]{revtex4-2} 
\usepackage[caption=false]{subfig}
\usepackage{graphicx}
\usepackage{dcolumn}
\usepackage{bm}
\usepackage{float}
\makeatletter
\let\newfloat\newfloat@ltx
\makeatother

\usepackage{algpseudocode}
\usepackage{algorithm}

\usepackage{float}
\usepackage[english]{babel}
\usepackage{tabularx}
\usepackage{hhline}
\usepackage{amsmath}
\usepackage[colorlinks=true, allcolors=blue]{hyperref}
\usepackage{tikz}
\usepackage{amssymb}
\usepackage{xcolor}
\usepackage{color}

\usetikzlibrary{shapes.geometric, arrows}
\tikzstyle{startstop} = [rectangle, rounded corners, minimum width=3cm, minimum height=1cm,text centered, draw=black, fill=red!30]
\tikzstyle{io} = [trapezium, trapezium left angle=70, trapezium right angle=110, minimum width=3cm, minimum height=1cm, text centered, draw=black, fill=blue!30]
\tikzstyle{process} = [rectangle, minimum width=3cm, minimum height=1cm, text centered, draw=black, fill=orange!30] 
\tikzstyle{decision} = [diamond, minimum width=3cm, minimum height=1cm, text centered, draw=black, fill=green!30]
\tikzstyle{arrow} = [thick,->,>=stealth]

\begin{document}

\title{Characterization of partial wetting by CMAS droplets using multiphase many-body dissipative particle dynamics and data-driven discovery  based on PINNs}

\author{Elham Kiyani}
\affiliation{Department of Mathematics, The University of Western Ontario, 1151 Richmond Street, London, Ontario, N6A~5B7, Canada}
\affiliation{The Centre for Advanced Materials and Biomaterials (CAMBR), The University of Western Ontario, 1151 Richmond Street, London, Ontario, N6A~5B7, Canada} 

\author{Mahdi Kooshkbaghi}
\affiliation{Simons Center for Quantitative Biology, Cold Spring Harbor Laboratory, Cold Spring Harbor, NY, USA}

\author{Khemraj Shukla}
\affiliation{Division of Applied Mathematics, Brown University, 182 George Street, Providence, RI 02912, USA}

\author{Rahul Babu Koneru}
\affiliation{Department of Aerospace Engineering,
University of Maryland, College Park, MD 20742, USA}

\author{Zhen Li}
\affiliation{Department of Mechanical Engineering,
Clemson University, Clemson, SC 29634, USA}

\author{Luis Bravo}
\affiliation{US Army Research Laboratory,
Aberdeen Proving Ground, MD 21005, USA}

\author{Anindya Ghoshal}
\affiliation{US Army Research Laboratory,
Aberdeen Proving Ground, MD 21005, USA}

\author{George Em Karniadakis}
\affiliation{Division of Applied Mathematics, Brown University, 182 George Street, Providence, RI 02912, USA}

\author{Mikko Karttunen}
\affiliation{The Centre for Advanced Materials and Biomaterials (CAMBR), The University of Western Ontario, 1151 Richmond Street, London, Ontario, N6A~5B7, Canada} 
\affiliation{Department of Physics and Astronomy,
 The University of Western Ontario, 1151 Richmond Street, London, Ontario,  N6A\,3K7, Canada}
\affiliation{Department of Chemistry, The University of Western Ontario, 1151 Richmond Street, London, Ontario, N6A~5B7, Canada}

\date{\today}
\begin{abstract}
The molten sand, a mixture of calcia, magnesia, alumina, and silicate, known as CMAS, is characterized by its high viscosity, density, and surface tension. The unique properties of CMAS make it a challenging material to deal with in high-temperature applications, requiring innovative solutions and materials to prevent its buildup and damage to critical equipment.  Here, we use multiphase many-body dissipative particle dynamics (mDPD) simulations to study the wetting dynamics of highly viscous molten CMAS droplets. The simulations are performed in three dimensions, with varying initial droplet sizes and equilibrium contact angles. We propose a coarse parametric ordinary differential equation (ODE) that captures the spreading radius behavior of the CMAS droplets. The ODE parameters are then identified based on the Physics-Informed Neural Network (PINN) framework. Subsequently, the closed form dependency of parameter values found by PINN on the initial radii and contact angles are given using symbolic regression. Finally, we employ Bayesian PINNs (B-PINNs) to assess and quantify the uncertainty associated with the discovered parameters. In brief, this study provides insight into spreading dynamics of CMAS droplets by fusing simple parametric ODE modeling and state-of-the-art machine learning techniques.
\end{abstract}

\maketitle

\section{Introduction}

Recent advancements in machine learning (ML) have opened the way for  
extracting governing equations directly from experimental (or other) data~\cite{brunton2016discovering,ren2020identifying,delahunt2022toolkit}.
One particularly exciting use of ML 
is the extraction of partial differential equations (PDEs) that describe the evolution and emergence of patterns or features~\cite{thiem2020emergent,kiyani2022machine,lee2020coarse,meidani2021data}.

Spreading of liquids on solid surfaces is a classic problem~\cite{De_Gennes1985-jy,Bonn2009-gg}. Although the theoretical foundations were laid by Young and Laplace already in the early 1800's~\cite{Young1805-sa,Laplace1805-ao}, there are still many open questions and it remains a highly active research field especially in the context of microfluidics~\cite{nishimoto2013bioinspired} as well as in the design of propulsion materials ~\cite{jain2021amr}. As discussed in detail in the review of  Popescu et al.~\cite{Popescu2012-db}, there are two fundamentally different cases: non-equilibrium spreading of the droplet, and the case of thermodynamic equilibrium when spreading has ceased and the system has reached its equilibrium state. 

\begin{figure*}[!tb]
    \centering
    \includegraphics[width=\textwidth]{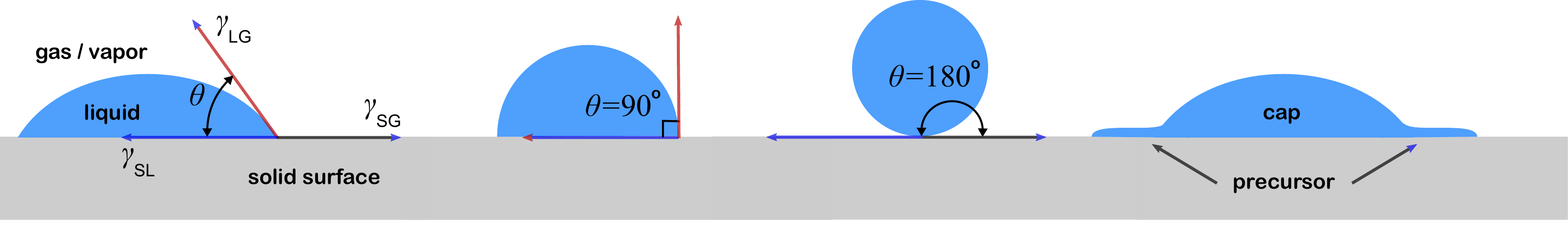}
    \caption{A schematic showing the equilibrium contact angle (that is, $\theta \equiv \theta_\mathrm{eq}$), the surface tensions ($\gamma$), the threshold between low- and high-wetting regimes ($\theta_\mathrm{eq} =90^{\circ}$),
    and a situation of a non-wetting droplet ($\theta_\mathrm{eq} =180^{\circ}$). The last panel demonstrates the occurrence of a precursor that is observed in some cases. In that case, the (macroscopic) contact angle is defined using the macroscopic part of the droplet as indicated by the black line in the rightmost figure. The height of the precursor is in the molecular length scales~\cite{Hardy1919-li,Nieminen1992-hg,Popescu2012-db}. 
    }
    \label{fig:contact_angle}
\end{figure*}

In thermodynamic equilibrium, the Laplace equation relates the respective surface tensions of the three interfaces  via~\cite{De_Gennes1985-jy,Popescu2012-db}
\begin{equation}
	\cos{\theta_\mathrm{eq}}= \frac{\gamma_{SG}-\gamma_{SL}}{\gamma_{LG}},
	\label{eq:young}
\end{equation}
where $\theta_\mathrm{eq}$ is the equilibrium contact angle, and $\gamma_{SG}$, $\gamma_{SL}$, and $\gamma_{LG}$ are the surface tensions between solid-gas, solid-liquid and liquid-gas phases, respectively (see Figure~\ref{fig:contact_angle}). Two limiting situations can be identified, namely, partial wetting and complete wetting. In the latter, the whole surface becomes covered by the fluid and $\theta_\mathrm{eq} = 0^{\circ}$, that is, $\gamma_{SG}\!- \! \gamma_{LG} \!-\!\gamma_{SL} = 0$. When the equilibrium situation corresponds to partial wetting, $\theta_\mathrm{eq} \ne 0^{\circ}$,  it is possible to identify the cases of
high-wetting ($0^{\circ} < \theta_\mathrm{eq} < 90^{\circ}$), low-wetting ($90^{\circ} \le \theta_\mathrm{eq} < 180^{\circ}$), and non-wetting ($\theta_\mathrm{eq} = 180^{\circ}$).

When a droplet spreads, it is out of equilibrium and properties such as viscosity and the associated processes need to be addressed~\cite{De_Gennes1985-jy,Bonn2009-gg,Popescu2012-db}. In experiments, the most common choice is to use high viscosity liquids in order to eliminate inertial effects. An early classic experiment by Dussan and Davis~\cite{dussan1979spreading} gave a beautiful demonstration of some of the phenomena. They added tiny drops of marker dye on the surface of a spreading liquid. They observed a caterpillar-type rolling motion of the marker on the surface giving rise to dissipation via viscous friction. Effects of viscosity and dissipation remain to be fully understood and they have a major role in wetting phenomena~\cite{cormier2012beyond,McGraw2016-ci,Edwards2020-in}.

Calcium-magnesium-aluminosilicate, CMAS, is a molten mixture of several oxides, including calcia (CaO), magnesia (MgO), alumina (Al$_2$O$_3$), and silicate (SiO$_2$).
It has a high melting point, typically
around 1,240$^\circ$C~\cite{Poerschke2015-gm} (although it can be significantly higher, see, e.g., Wiesner et al. and references therein~\cite{Wiesner2016-pj}),
which allows it to exist in the molten state even at high temperatures encountered in modern aviation gas turbine engines~\cite{Clarke2012-um,Ndamka2016-tx}.
With high viscosity, high density, 
and high surface tension, CMAS tends to form non-volatile droplets of $\theta_\mathrm{eq}\ne 0$ rather than completely wetting the surfaces~\cite{nieto2021calcia,grant2007cmas,VIDALSETIF201239}.
When it solidifies,
it forms a glass-like material that can adhere to surfaces and resist erosion. The buildup of CMAS on turbine engine components can lead to clogging of the cooling passages and degradation of the protective coatings, resulting in engine performance issues and even damage or failure~\cite{Ndamka2016-tx,Song2016-no, Wiesner2016-pj,Clarke2012-um}.

The spreading of a droplet over a solid surface is commonly characterized using a power law, $r \!\sim \! t^{\alpha}$, which expresses the radius of the wetted area as a function of time. The relationship is called Tanner's law for macroscopic completely wetting liquids at late times with $\alpha = 1/10$~\cite{Tanner1979-mp,Bonn2009-gg}. 
Power laws have also been demonstrated at microscopic scales~\cite{Nieminen1992-hg}.
However, several conditions such as surface properties, droplet
shape, and
partial wetting result in deviations from Tanner's law~\cite{mchale2004topography,cormier2012beyond,Winkels2012-aw}.

A common method, and as the above suggests,  for analyzing the spreading dynamics is to investigate existence of the power law behavior. 
To determine the presence of such power-law regimes in data, one can simply employ
\begin{equation}
    \alpha (t) =\frac{d \ln (r)}{d \ln (t)} .
\label{eq:alpha}
\end{equation}
While this has worked remarkably well for complete wetting by viscous fluids, the situation for partial wetting is different~\cite{Winkels2012-aw}. 
In our study, we investigate the spreading behavior of CMAS droplets using
multiphase many-body dissipative particle dynamics (mDPD) simulations. We generalize Equation~\eqref{eq:alpha} such that it includes dependence on the initial droplet radius $R_0$ and $\theta_\mathrm{eq}$ in order to describe partial wetting, that is, $\alpha \equiv \alpha(t, R_0 , \theta_\mathrm{eq})$.  

Our objective is to gain a comprehensive understanding of the behavior of CMAS droplet spreading dynamics by integrating knowledge about the fundamental physics of the system into the neural network architecture. To achieve this, we employ the framework of Physics-Informed Neural Networks (PINNs)~\cite{raissi2019physics}, an emerging ML technique that incorporates the physics of a system into deep learning.
PINNs address the challenge of accurate predictions in complex systems with varying initial and boundary conditions. By directly incorporating physics-based constraints into the loss function, PINNs enable the network to learn and satisfy the governing equations of the system.

The ability of PINNs to discover equations makes them promising for applications in scientific discovery, engineering design, and data-driven modeling of complex physical systems~\cite{karniadakis2021physics}.
Their integration of physics-based constraints into the learning process enhances their capacity to generalize and capture the underlying physics accurately.
Here, we also employ symbolic regression to generate a mathematical expression for each unknown parameter. 
Furthermore, we employ Bayesian Physics-Informed Neural Networks (B-PINNs)~\cite{yang2021b} to quantify the uncertainty of the predictions.

The rest of this article is structured as follows: In Section~\ref{mDPD}, we provide an overview of mDPD simulation parameters and system setup. Simulation outcomes and the data preparation process are presented in Section~\ref{simulation-results}. Section~\ref{PINNs} gives a brief introduction to the PINNs architecture, followed by a presentation of the results of PINNs and parameter discovery. The symbolic regression results and the mathematical formulas for the parameters are presented in Section~\ref{Symbolic_regression}. Section~\ref{B-PINN} covers the discussion on B-PINNs as well as the quantification of uncertainty in predicting the parameters.
Finally, we conclude with a summary of our work in Section~\ref{Conclusion}.

\section{Multiphase many-body dissipative particle dynamics simulations}\label{mDPD}

Three-dimensional simulations were performed using the 
mDPD
method~\cite{rao2021modified,xia2017many,li2013three}, which is an extension of the traditional dissipative 
particle dynamics (DPD) model~\cite{Espanol1995-wg,Groot2004-yj}. DPD is a mesoscale simulation technique for studies of complex fluids, particularly multiphase systems, such as emulsions, suspensions, and polymer blends~\cite{zhao2021review,lei2018many,ghoufi2012coarse}. The relation between DPD and other coarse-grained methods and atomistic simulations have been studied and discussed by Murtola et al.~\cite{Murtola2009-gr}, Li et al.~\cite{Li2016AComparative, Chan2023AMori}, and Espa{\~n}ol and Warren~\cite{Espanol2017Perspective}. 

In DPD and mDPD models, the position ($\vec{r}_i$) and velocity ($\vec{v}_i$) of a particle $i$ with a mass $m_i$ are governed by Newton's equations of motion in the form of
\begin{eqnarray}
\frac{d\vec{r}_i}{dt} & = & \vec{v}_i, \\ \nonumber
m_i \frac{d\vec{v}_i}{dt} & = & \vec{F}_{i}  =  \sum_{j\ne i}\vec{F}_{ij}^\mathrm{C} + \vec{F}_{ij}^\mathrm{D} + \vec{F}_{ij}^\mathrm{R}. 
\label{eq:sum_of_forces}
\end{eqnarray}
The total force on particle $i$, that is, $\vec{F}_i$, consists of three pairwise components, i.e., the conservative $\vec{F}^\mathrm{C}$, dissipative $\vec{F}^\mathrm{D}$, and random forces $\vec{F}^\mathrm{R}$. The latter two are identical in DPD and mDPD models, given by,
 \begin{eqnarray}
\vec{F}_{ij}^\mathrm{D} & = & - \gamma \omega_\mathrm{D}(r_{ij}) ( \vec{v}_{ij} \cdot  \vec{e}_{ij}) \vec{e}_{ij},\\
\vec{F}_{ij}^\mathrm{R} & = & \zeta \omega_\mathrm{R}(r_{ij}) (dt)^{-1/2} \xi_{ij} \vec{e}_{ij},
\label{eq:thermostat}
\end{eqnarray}
where $\vec{e}_{ij}$ is a unit vector, $\omega_{D}$ and  $\omega_{R}$ are weight functions for the dissipative and random forces, and $\xi_{ij}$ a pairwise conserved Gaussian random variable with zero mean and second moment $\langle \xi_{ij}(t) \xi_{kl} (t') \rangle = ( \delta_{ik} \delta_{jl} + \delta_{il} \delta_{jk}) \delta(t-t')$, where $\delta_{ij}$ is the Kronecker delta and $\delta (t-t')$ the Dirac delta function. Together, the dissipative and random forces constitute a momentum conserving Langevin-type thermostat. 
The weight functions and the constants $\gamma$ and $\zeta$ are related via fluctuation-dissipation relations first derived by Espa\~{n}ol and Warren~\cite{Espanol1995-wg}
\begin{eqnarray}
\omega_{D} & = & \left(\omega_{R} \right)^2, \\
\zeta & = & \sqrt{2\gamma k_\mathrm{B}T},
\label{eq:FD}
\end{eqnarray}
in which $k_\mathrm{B}$ is the Boltzmann constant and $T$ the temperature. This relation guarantees the canonical distribution~\cite{Espanol1995-wg} for fluid systems in thermal equilibrium. The functional form of the weight function is not specified, but the most common choice (also used here) is 
\begin{equation}
  \omega_\mathrm{D}(r_{ij}) = \left\{
  \begin{array}{ll}
    \left(1 - r_{ij}/r_d\right)^s & \mbox{for $r_{ij} \le r_\mathrm{d}$} \\
    0 & \mbox{for $r_{ij}>r_\mathrm{d}$}.
\end{array}
\right.
\label{eq:omega}
\end{equation}
Here, 
$s=1.0$ is used and $r_\mathrm{d}$ defines a cutoff distance for the dissipative and random forces. 

Although the above equations are the same for both DPD and mDPD, they differ in their conservative forces. Here, we use the form introduced by Warren~\cite{Warren2001-qm,Warren2003-qp},
\begin{equation}
F^\mathrm{C}_{ij} = A \omega_\mathrm{C}(r_{ij})\vec{e}_{ij} + B (\rho_i + \rho_j) \omega_\mathrm{B} \vec{e}_{ij}.   
\label{eq:mDPD}
\end{equation}
The functional form of both weight functions $\omega_C$ and $\omega_B$ is the same as $\omega_\mathrm{D}$ in Equation~\eqref{eq:omega} but with different cutoff distances $r_c$ and $r_b$. The first term in Equation~\eqref{eq:mDPD} is the standard expression for the conservative force in DPD, and the second one is the multi-body term. The constants $A$
and $B$ are chosen such that $A <0$ for attractive interactions and $B>0$ for repulsive interactions; note that in conventional DPD $A >0$ and $B=0$. The key component is the weighted local density
\begin{equation}
\rho_i = \sum_{j \ne i} \omega_\rho(r_{ij}).
\label{eq:w_density}
\end{equation}
There are several ways to choose the weight function~\cite{zhao2021review} and here, the normalized Lucy kernel~\cite{Lucy1977-mt} in 3-dimension is used,
\begin{equation}
 \omega_\rho(r_{ij}) = \frac{105}{16 \pi r^3_{\mathrm{c}\rho}}
 \left(
1+ \frac{3 r_{ij}}{r_{\mathrm{c}\rho}}
 \right)
 \left(
1- \frac{r_{ij}}{r_{\mathrm{c}\rho}}
 \right)^3,
\label{eq:lucy}
\end{equation}
with a cutoff distance $r_{c\rho}$ beyond which the weight function $\omega_\rho$ becomes zero.

\subsection{Simulation parameters and system setup}

To simulate molten CMAS, the parameter mapping of Koneru et al.~\cite{koneru2022quantifying} was used together with the open-source code LAMMPS~\cite{thompson2022lammps}.
In brief, the properties of molten CMAS at about 1,260$^\circ$\,C  based on the experimental data from Naraparaju et al.~\cite{Naraparaju2019-vh}, Bansal and Choi~\cite{Bansal2014-gn}, and Wiesner et al.~\cite{Wiesner2016-pj}, were used. In physical units, density was 2,690\,kg/m$^3$, surface tension 0.46\,N/m, and viscosity 3.6\,Pa$\cdot$s. Using the density and surface tension to estimate the capillary length ($\kappa = (\sigma /(\rho g))^{1/2}$) gives 4.18\,mm. The droplets in the simulations (details below) had linear sizes shorter than the capillary length and hence gravity was omitted. In terms of physical units, time: $6.297\!\times \!10^{-6}$\,s, length: $17.017\!\times\! 10^{-6}$\,m, mass: $1.964 \! \times \! 10^{-8}$\,kg.

\begin{figure}
\centering
\includegraphics[width=0.49\textwidth]{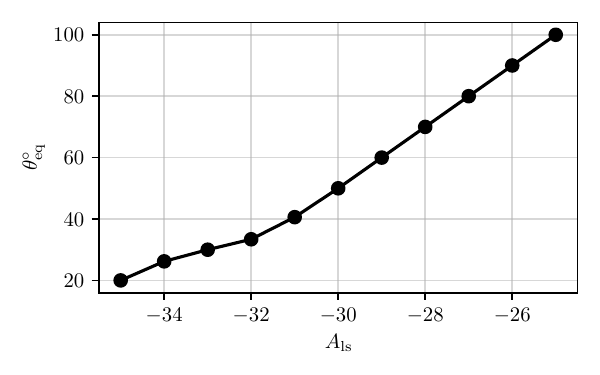}
\caption {The equilibrium contact angles $\theta_\mathrm{eq}$ for the different attraction parameters between the liquid and solid particles ($A_\mathrm{ls}$; see Equation~\eqref{eq:mDPD}). It is worth noting that the data for this figure has been extracted from  Koneru et al.~\cite{koneru2022quantifying}.}
\label{fig:contact_angles}
\end{figure}

Using the above values, droplets of initial radii of $d = 8,  9 , 10, 11$, and $12$ in mDPD units, 
corresponding to $ R_{0}= 0.136$\,mm, $0.153$\,mm, $0.17$\,mm, $0.187$\,mm, and $0.204$\,mm, respectively, were used in the simulations;  all of them are smaller than the capillary length. The time step was 0.002 (mDPD units) corresponding to 12.59\,ns. In addition, $k_\mathrm{B}T \!= \!1$, $r_c \!=\!1.0$, $r_b=r_{\mathrm{c}\rho}\!=\!0.75$, $r_d=1.45$, $\gamma \!= \!20$, and $B\!=\!25$.  
The attraction parameter, $A$ in Equation~\eqref{eq:mDPD} has to be set for the interactions between the liquid particles ($A_\mathrm{ll}$), and the liquid and solid particles ($A_\mathrm{ls}$). The former was set to $A_\mathrm{ll}=-40$ and $A_\mathrm{ls}$ was chosen based on simulations that provided the desired $\theta_\mathrm{eq}$, thus allowing for controlled variation of $\theta_\mathrm{eq}$ (see Figure~\ref{fig:contact_angles}).

The initial configuration of the droplet and the solid wall were generated from a random distribution of equilibrated particles with a number density $\rho=6.74$. This amounts to about 60,660 particles in the wall and depending on the initial radius, anywhere between 14,456 and 48,786 particles in the droplet. Periodic boundary conditions are imposed along the lateral directions and a fixed, non-periodic boundary condition is imposed along the wall-normal direction. Since mDPD is a particle-based method, the spreading radius and the dynamic contact angle are approximated using surface-fitting techniques. First, the outermost surface of the droplet is identified based on the local number density, i.e., particles with $\rho \in [0.45, 0.6]$. The liquid particles closest to the wall are fitted to a circle of radius $r$, i.e., the spreading radius. On the other hand, a sphere with the centroid of the droplet as the center is fit to the surface particles to compute the contact angle. The contact angle is defined as the angle between the tangent at the triple point (liquid-solid-gas interface) and the horizontal wall. The wall in these simulations is made-up of randomly distributed particles to eliminate density and temperature fluctuations at the surface. 
Following~\cite{li2018dissipative}, the root mean squared height ($R_q$) of the surface scales linearly with $1/\sqrt{N_w}$ where $N_w=2\pi r_{cw}^3/3 \cdot \rho_w$ is the number of neighboring particles. In this work, $R_q$ comes out to be around 0.0708 mDPD units or 1.2\,$\mu$m.

\begin{figure}[!tb]
    \centering
    \includegraphics[trim=7 0 0 0,clip,width=0.27\textwidth]{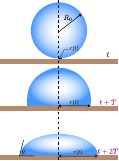}
    \includegraphics[trim=7 0 0 0,clip,width=0.20\textwidth]{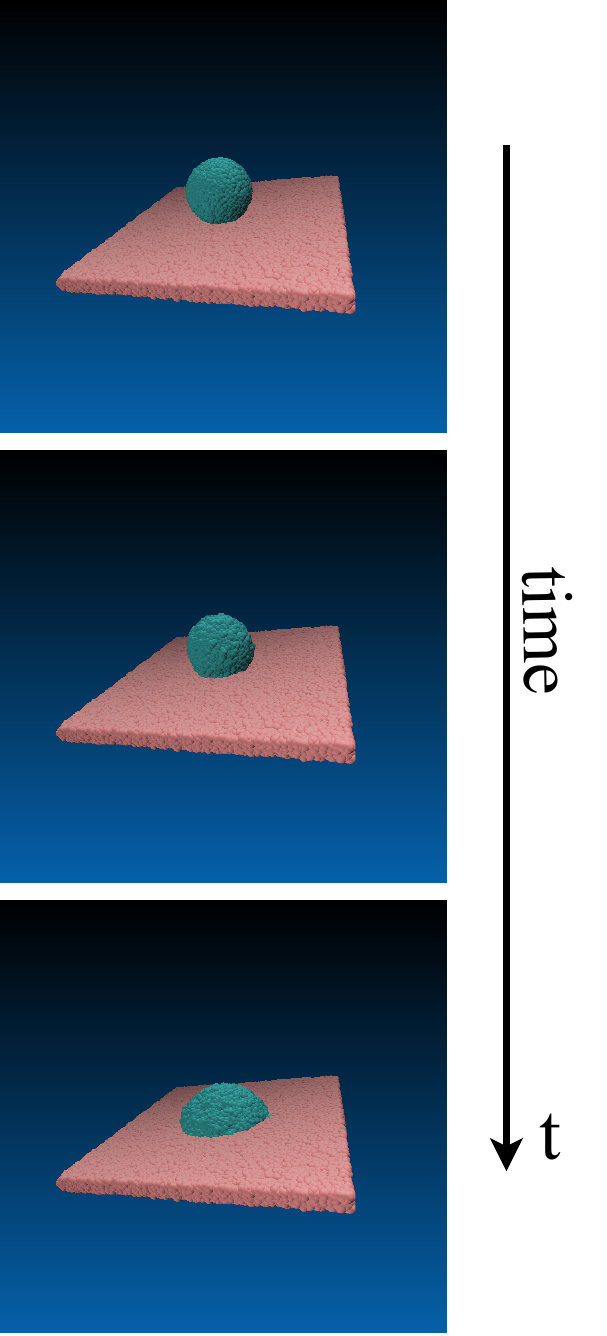}
    \caption{Left: Illustration of the spreading behavior of a CMAS droplet on a high surface energy surface at  different times. The droplet with initial size of $R_0$ spreads on the surface with radius $r(t)$ and contact angle $\theta (t)$. Right: A series of snapshots from a simulation of a droplet with initial size of $R_0 = 0.136$\,mm and an equilibrium contact angle of $\theta_\mathrm{eq} = 93.4^\circ$.
    }
    \label{fig:drop}
\end{figure}

As the CMAS droplet spreads on the substrate, it loses its initial spherical shape and begins to wet the surface as depicted in Figure~\ref{fig:drop}, forming a liquid film between the droplet and the substrate. 
Understanding how droplets behave on surfaces is important for a wide range of applications, including in industrial processes, microfluidics, propulsion materials, and the design of self-cleaning surfaces~\cite{pitois1999crystallization,chen2016droplet,hassan2019self,jain2021amr,nieto2021calcia}.

\section{Simulation results}\label{simulation-results}

\begin{figure}
\centering
\includegraphics[width=0.49\textwidth]{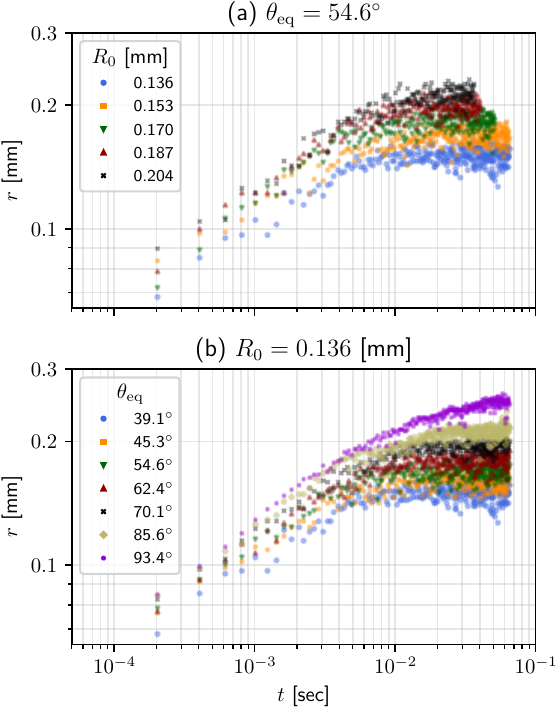}
\caption {The impact of $\theta_\mathrm{eq}$ and $R_0$ on the droplet radii as a function of time for various
(a) initial drop sizes with equilibrium contact angle of $\theta_\mathrm{eq} = 54.6^\circ$ corresponding to $A_\mathrm{ls} =  30.0$, and (b) equilibrium contact angles (corresponding to $A_\mathrm{ls} = -25.0, -25.8, -27.0, -28.0, -29.0, -30.0, -31.4, -32.2$) and initial drop size $R_{0}=0.136$\,mm.
}
\label{fig:radius}
\end{figure}

The size of a droplet changes over time.  By tracking the changes, we can gain insight into the physical
processes involved in spreading.
The time evolution of the droplet radius ($r(t)$) 
is shown in Figure~\ref{fig:radius}. The log-log plots show the effect of the initial drop size $R_0$ and equilibrium contact angles $\theta_\mathrm{eq}$ on the radius $r(t)$. 
Figure~\ref{fig:radius}(a) displays $r(t)$ for initial drop sizes $R_0$ of $0.136$\,mm, $0.153$\,mm, $0.17$\,mm, $0.183$\,mm, and $0.204$\,mm and equilibrium contact angle of $\theta_\mathrm{eq}=54.6^\circ$. Similarly, Figure~\ref{fig:radius}(b) shows the spreading radius for different equilibrium contact angles ($\theta_\mathrm{eq}=93.4^\circ$, $85.6^\circ$, $77.9^\circ$, $70.1^\circ$, $62.4^\circ$, $54.6^\circ$, $45.3^\circ$, and $39.1^\circ$) with an initial drop size of $R_0=0.136$\,mm.

Eddi et al.~\cite{eddi2013short}
used high-speed imaging with time resolution covering six decades to study the spreading of water-glycerine mixtures on glass surfaces. By varying the amount of glycerine, they were able to vary the viscosity over the range 0.0115-1.120\,Pa$\cdot$s. They observed two regimes, the first one for early times with $\alpha$ changing continuously as a function of time from  $\alpha \approx 0.8$ to $\alpha \approx 0.5$. This was followed by a sudden change to the second regime in which $\alpha$ settled to $0.1 < \alpha < 0.2$. As pointed out by Eddi et al.~\cite{eddi2013short}, the second regime agrees with Tanner's law~\cite{Tanner1979-mp}. All of their systems displayed complete wetting.

\begin{figure}   
\centering
\includegraphics[width=0.49\textwidth]{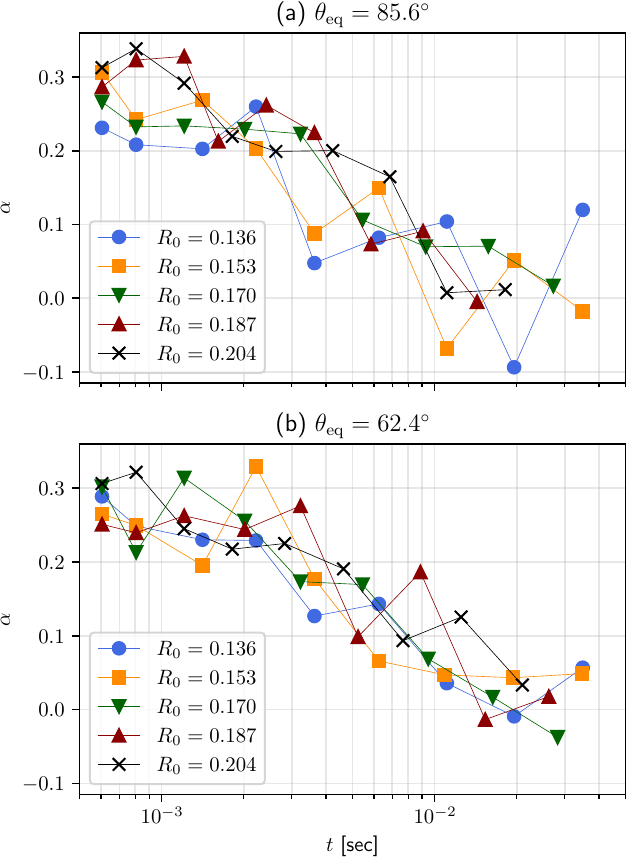}
\caption { The value of $\alpha$, calculated using Equation~\eqref{eq:alpha}, varies for different initial radii and fixed equilibrium contact angles $\theta_\mathrm{eq}=62.4^\circ$ and $\theta_\mathrm{eq}=85.6^\circ$. The figure illustrates that $\alpha$ is influenced by both the initial drop size $R_0$ and the equlibrium contact angle $\theta_\mathrm{eq}$.}
\label{fig:simulation-alpha}
\end{figure}

Based on $r(t)$ 
of the CMAS drops and $\alpha$ shown in Figure~\ref{fig:radius}, it can be observed that
$r(t)$ (and based on the power-law $\alpha$) depends both on the initial drop size and the equilibrium contact angle.
The values of $\alpha$ for some 
simulation datasets are plotted over time in Figure~\ref{fig:simulation-alpha}. The plot shows the behaviour of $\alpha$ for different initial drop sizes and equilibrium contact angles of $\theta_\mathrm{eq}=62.4^\circ$  and $\theta_\mathrm{eq}=85.6^\circ$.

Inspired by the experimental results of Eddi et al.~\cite{eddi2013short}, the simulations of Koneru et al.~\cite{koneru2022quantifying}, and the current simulations,
we propose a simple sigmoid type dependence for $\alpha$,
\begin{equation} 
\label{ODE}    
\!\!\!\!   
\frac{d\ln (r)}{d\ln (t)} \! = \!\alpha(t, R_0 , \theta_\mathrm{eq}) \! := \! \eta \! \left[\frac{1}{1 \! + \!
    \exp\left(\beta (\tau \!-\! \ln(t)\right) } \! - \! 1\right]\!.
\end{equation}
The two constant values of $\alpha$ discussed in the above references are the two extrema of the sigmoid curve, given that the transition between the two regimes occurs at $\ln(t_\mathrm{transition})=\tau$. 
The parameters of Equation~\eqref{ODE} are discovered by PINNs and their dependence on $R_0$ and $\theta_\mathrm{eq}$ is then expressed using symbolic regression.
\begin{figure*}[!htp]
\centering
\includegraphics[width=\textwidth]{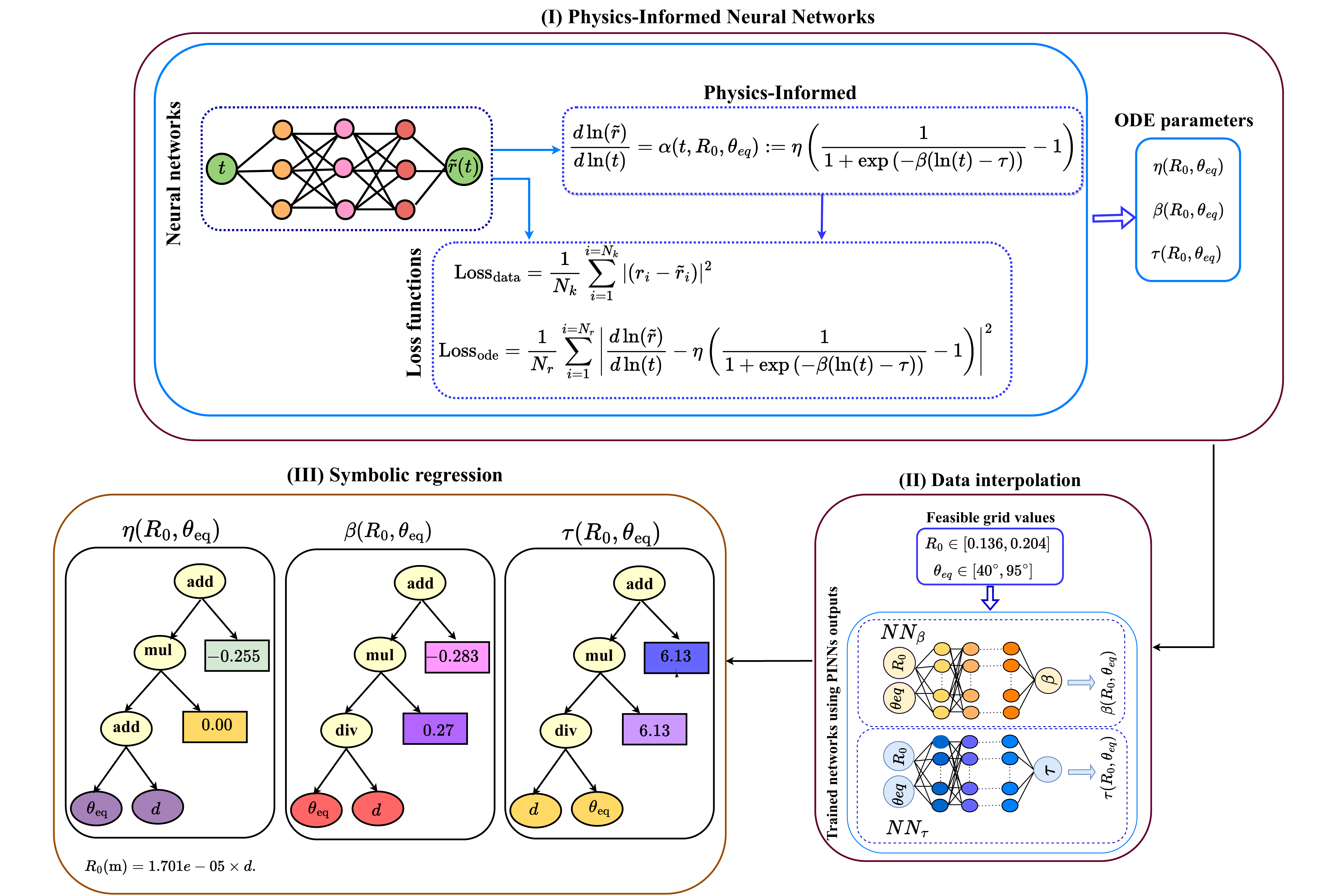}
\caption{The process of utilizing PINNs to extract three unknown parameters of the ODE~\eqref{ODE}, using three-dimensional mDPD simulation data. First, a neural network is trained using simulation data, where the input is time $t$ and the output is spreading radii $\tilde{r}(t)$. This neural network comprises four layers with three neurons and is trained for $12,000$ epochs. Subsequently, the predicted $\tilde{r}(t)$ is used to satisfy Equation~\eqref{ODE} in the physics-informed part. The loss function for this process consists of two parts: data matching and residual. By optimizing the loss function, the values of $\eta(R_0, \theta_\mathrm{eq})$, $\beta(R_0, \theta_\mathrm{eq})$, and $\tau(R_0,\theta_\mathrm{eq})$ are determined for each set of $R_0$ and $\theta_\mathrm{eq}$. After predicting the unknown parameters using PINNs, two additional neural networks, denoted as $NN_\beta$ and $NN_\tau$, are trained using these parameters to generate values for the unknown parameters at points where data is not available. The outputs of these networks, together with the outputs of the PINNs, are then fed through a symbolic regression model to discover a mathematical expression for discovered parameter.
}
\label{fig:full-steps}
\end{figure*} 
The general steps in the discovery of the droplet spreading equation and the extraction of the unknown parameters $\eta(R_0, \theta_\mathrm{eq})$, $\beta(R_0, \theta_\mathrm{eq})$, and $\tau( R_0, \theta_\mathrm{eq})$ are shown in Figure~\ref{fig:full-steps} and can be summarized as follows:
\begin{itemize}
    \item Data collection: For this study, data is collected by conducting three-dimensional simulations using the mDPD method in LAMMPS with varying initial drop sizes $R_0$ and equilibrium contact angles $\theta_\mathrm{eq}$. 
    \item PINNs: The input of the network is time $t$ and output of the network is the spreading radii $\tilde{r}(t)$. The physics informed part of PINNs encapsulated in designing the loss function. In this study, the ``goodness'' of the fit is measured by (a) deviation from trained data together with (b) deviation of network predictions and those from ODE \eqref{ODE} solutions. This optimization process reveals the the unknown parameters $\eta( R_0, \theta_\mathrm{eq})$, $\beta( R_0, \theta_\mathrm{eq})$, and $\tau( R_0, \theta_\mathrm{eq})$.
    \item Data interpolation: After PINNs are trained, we used their predictions together with two additional multilayer perceptron neural networks to fill the sparse parameter space. This step helps our next goal which is relating the ODE parameters to $R_0$ and $\theta_\mathrm{eq}$ without performing three-dimensional simulations.
    \item Symbolic regression: Discovering a mathematical expression for each unknown parameter, $\eta(R_0, \theta_\mathrm{eq})$, $\beta(R_0, \theta_\mathrm{eq})$, and $\tau( R_0, \theta_\mathrm{eq})$ of the Equation~\eqref{ODE}.

    \item In order to quantify the uncertainty associated with our predictions, we utilize B-PINN and leverage the insights gained from PINNs' prediction and symbolic regression, with a specific emphasis on the known value of $\eta$.  By employing B-PINN, we can effectively ascertain the values of two specific parameters, $\beta$ and $\tau$, which in turn enable us to quantify the uncertainty in our predictions.

\end{itemize}

\section{Physics-Informed Neural Networks (PINNs)}\label{PINNs}

PINNs are a promising approach that leverages the flexibility and scalability of deep neural networks to solve or even to discover governing equations, incorporating physical laws and constraints into the network structure~\cite{raissi2019physics, karniadakis2021physics,shukla2020physics-old,mishra2020estimates,chen2020physics,MAO2020112789}. 

The training phase of PINNs consists of an optimization process applied on top of a neural network structure to identify
the set of parameters in the governing equations with a pre-defined form (PDEs or ODEs). Those parameters aim to satisfy both data and the physical constraints.

\subsection{Discovering parameters of ODE}

As discussed in Section~\ref{simulation-results}, our study aims to identify the values of the parameters $\eta(R_0,\theta_\mathrm{eq})$, $\beta(R_0,\theta_\mathrm{eq})$, and $\tau(R_0, \theta_\mathrm{eq})$ in the ODE given by Equation~\eqref{ODE}. 
The general steps of the framework are shown in Figure~\ref{fig:full-steps}(I). 
PINNs take time $t$ as an input to predict $\tilde{r}(t)$ at each time. The network architecture consists of four layers with three neurons each and is trained for $12,000$ epochs with
the learning rate of $10^{-3}$.
\begin{figure*}[!htp]
\centering
\includegraphics[width=0.99\textwidth]{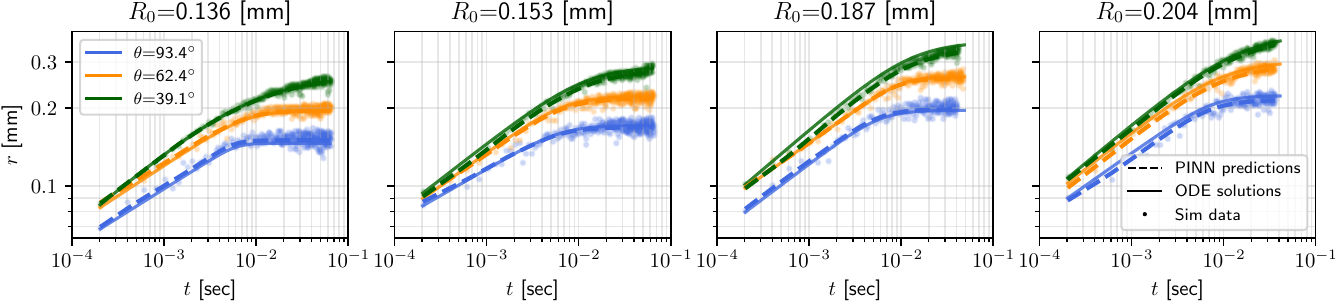}
\caption{
Comparison of the time evolution of the droplet radii: 
mDPD simulations (symbols), ODE model~\eqref{ODE} (solid lines) and PINN predictions (dashed lines) for $\theta_\text{eq}=\{39.1^\circ, 62.4^\circ,93.4^\circ\}$ and  $R_0=\{0.136, 0.153, 0.187,0.204\}$\,mm parameter sets.
}
\label{fig:predicted-r(t)}
\end{figure*} 

\begin{figure*}
\centering
\includegraphics[width=0.99\textwidth]{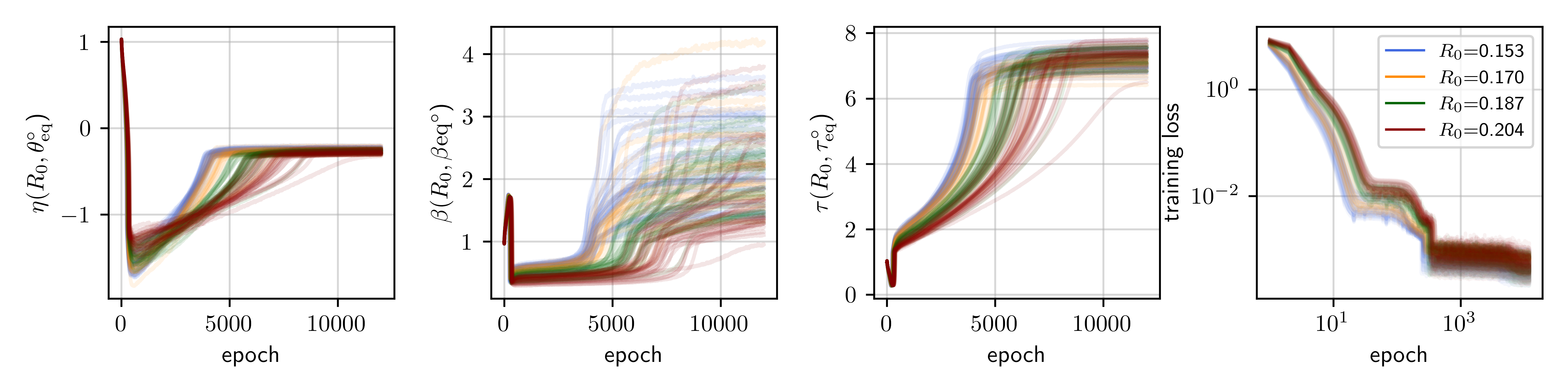}
\caption{ The first three plots show the evolution of parameters $\eta(R_0,\theta_\mathrm{eq})$, $\beta(R_0,\theta_\mathrm{eq})$, and $\tau(R_0,\theta_\mathrm{eq})$ over multiple epochs. These plots demonstrate that the parameters gradually converge to a stable state after $12,000$ epochs. The rightmost figure displays the traces of the loss function for the PINNs framework. The learning curves demonstrate the decreasing trend of the loss functions, indicating that they converge to a stable point for all initial drop sizes and $\theta_\mathrm{eq}$. }
\label{fig:loss}
\end{figure*}

\begin{figure*}
\centering
\includegraphics[width=\textwidth]{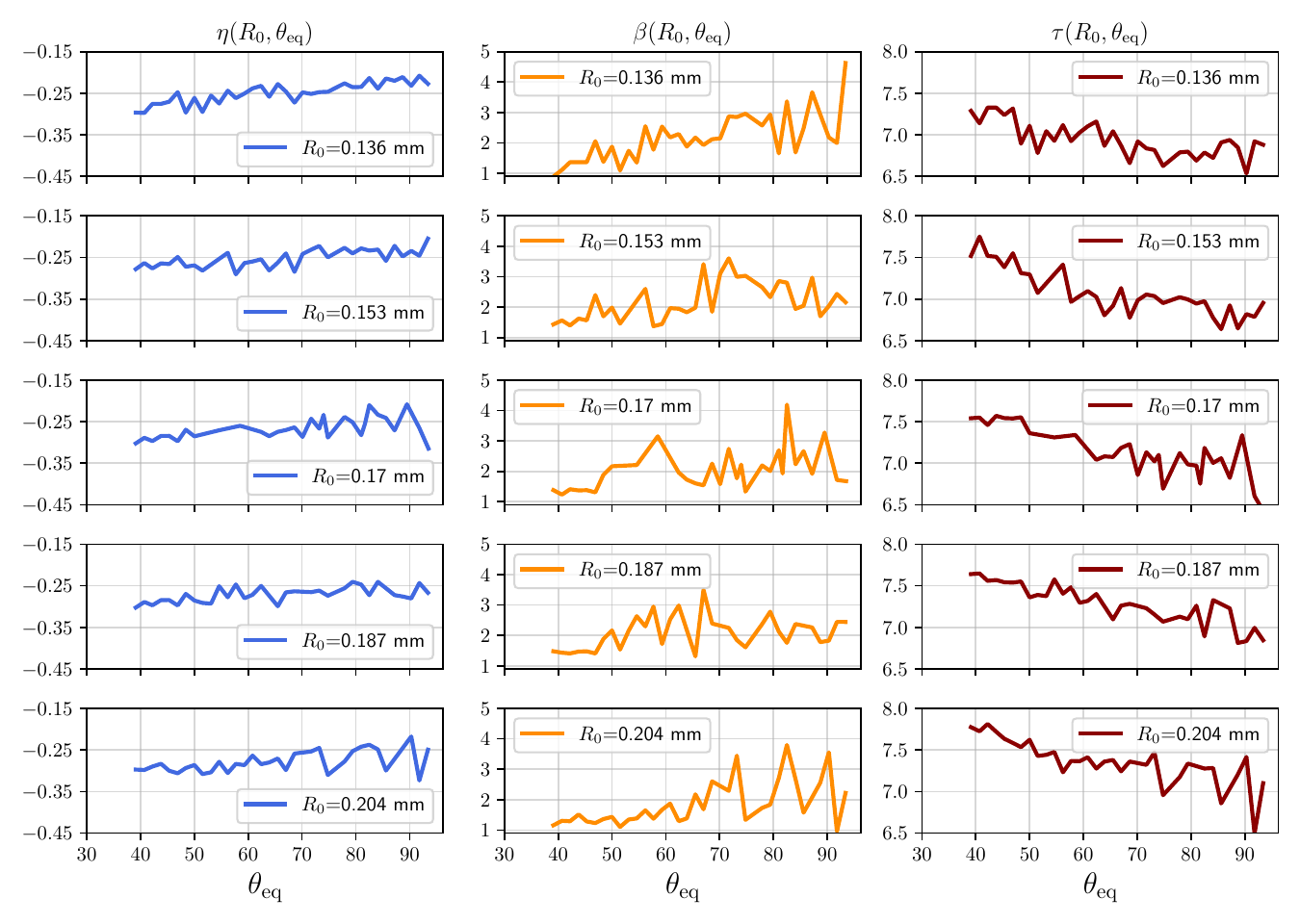}
\caption{The values of $\eta$, $\beta$, and  $\tau$ obtained through PINNs. These values exhibit varying behavior depending on the initial radius $R_{0}$ and equilibrium contact angles $\theta_\mathrm{eq}$. The horizontal axes display  the equilibrium contact angles $\theta_\mathrm{eq}$. The vertical axes of all figures represent the values of $\eta$ and $\beta$, and $\tau$. $\eta$ remains nearly constant within a small range of values between $-0.325$ and $-0.200$ and $\beta$ as well as $\tau$ change within a range of $1.0$ to $5.0$ and $6.5$ to $8.0$, respectively.}
\label{fig:abt0}
\end{figure*}

The 
predicted values for the time evolution of the radii $\tilde{r}(t)$ should satisfy the data and the physics-informed step, i.e., meet the requirements of the ODE, Equation~\eqref{ODE}. The two-component loss function, designed to meet the requirements, consists of $\mathrm{Loss_{data}}$ 
and $\mathrm{Loss_{ODE}}$,
\begin{subequations}
\label{eq:loss}
\begin{align}
  & \mathrm{Loss_{data}}=\frac{1}{N_{k}} \sum_{i=1}^{i=N_{k}} | (r_{i}(t) - \tilde{r}_{i}(t)) | ^{2}, \\ 
  & \mathrm{Loss_{ODE}} = 
  \frac{1}{N_{r}} \sum_{i=1}^{i=N_{r}} \left | \frac{d\ln (\tilde{r}) }{d\ln (t)} \right. - \nonumber \\ 
  &  \left .
  \eta \left(\frac{1}{1+\exp\left(-\beta(\ln(t)-\tau)\right)}-1\right) \right | ^{2},
\end{align}
\end{subequations}
where $\tilde{r}(t)$ and $r(t)$ stand for the radii from the prediction and simulation, respectively. $N_{k}$ is the number of training points and $N_{r}$ is the number of residual points. 
Figures~\ref{fig:predicted-r(t)}, \ref{fig:loss}, and \ref{fig:abt0} illustrate the results obtained by utilizing PINNs to discover the parameters of the ODE (Equation~\eqref{ODE}), which describes the dynamics of the 
radii of the CMAS drops.

Figure~\ref{fig:predicted-r(t)} shows comparisons of the simulation data ($r(t)$), the prediction ($\tilde{r}(t)$) and solution of the ODE, Equation~\eqref{ODE}.
The figure
demonstrates remarkable degree of agreement between simulations, PINNs and our ODE model.

The first three panels in Figure~\ref{fig:loss} show the convergence of $\eta(R_0,\theta_\mathrm{eq})$, $\beta(R_0,\theta_\mathrm{eq})$, and $\tau(R_0,\theta_\mathrm{eq})$ parameters during training. 
One can conclude that, all three parameters stabilize roughly after $10,000$ epochs.
In the rightmost panel of  Figure~\ref{fig:loss}, the loss function (Equation~\eqref{eq:loss}) history is plotted against the training epochs, stabilizing around $10^{-4}$. This indicates successful training of the PINNs model. The results shown in Figures~\ref{fig:predicted-r(t)}~and~\ref{fig:loss} demonstrate the capability of our proposed framework to accurately predict the spreading radius of CMAS across the different initial radii and equilibrium contact angles.

Each column in Figure~\ref{fig:abt0} shows
the values of
$\eta(R_0,\theta_\mathrm{eq})$, 
$\beta(R_0, \theta_\mathrm{eq})$, and 
$\tau(R_0,\theta_\mathrm{eq})$ obtained by PINNs for different initial radii $R_{0}$ ranging from $0.136$ to $0.204$\,mm and equilibrium contact angles $\theta_\mathrm{eq}$ ranging from $39.1^\circ$ to $93.4^\circ$.
The results show that $\eta$ changes within a small window between $-0.325$ and $-0.200$ for all $R_{0}$ and $\theta_\mathrm{eq}$. 
However, the changes in $\beta$ (between $1$ and $5$) and $\eta$ (between $6.0$ and $8.0$) are significant, indicating that these parameters are strongly depend on $R_{0}$ and $\theta_\mathrm{eq}$.

\subsection{Generate more samples of feasible radii and contact angles}\label{grid_points}

As discussed earlier, the parameters in our ODE model (Equation~\eqref{ODE}) are functions of the initial radius and the equilibrium contact angle.
Using PINNs, we were able to find the 
values for those parameters.
To find a closed-form relation between the parameters, $R_0$, and $\theta_\mathrm{eq}$, more data than the rather small current set is needed.
Performing three-dimensional mDPD simulations are, however, computationally expensive. 
In this section, we train two additional neural networks to capture the nonlinear relation between the ODE parameters found by PINNs, and the variables $R_0$ and for $\theta_{\mathrm{eq}}$. Then, we will use these trained networks to fill our sparse parameter space to perform symbolic regression in the next section.

Specifically, we generate values for $R_{0}$ in the interval $[0.136\,\mathrm{mm}, 0.204\,\mathrm{mm}]$ and $\theta_\mathrm{eq}$ in the range $[40^\circ, 95^\circ]$, as shown in Figure~\ref{fig:full-steps}(II).
Two fully connected networks, $NN_{\beta}$ and $NN_{\tau}$ consist of eight dense layers with $256/256/256/128/64/32/16/8$ neurons.
These networks are trained using an Adam optimizer with a learning rate of $10^{-2}$ for a total of $4,000$ epochs.

\begin{figure}
\centering
\begin{tabular}{c}
(a) \\
\includegraphics[width=0.49\textwidth]{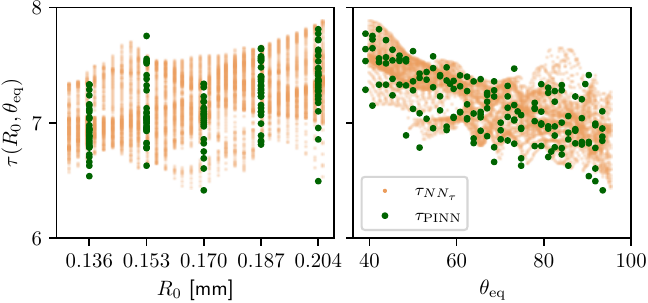} \\
(b)\\
\includegraphics[width=0.49\textwidth]{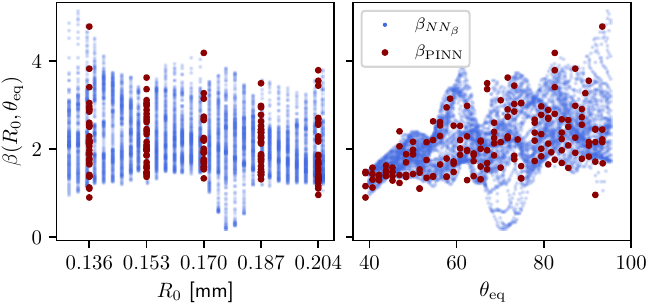}
\end{tabular}
\caption{Predictions of the parameters (a) $\tau$ and (b) $\beta$ using the trained neural networks $NN_{\beta}$ and $NN_{\tau}$. The horizontal axes show 
$R_0$ and $\theta_\mathrm{eq}$. 
The green and red circles correspond to the obtained values of $\tau$ and $\beta$ using the PINNs that were used to train $NN_{\beta}$ and $NN_{\tau}$. Additionally, the orange and blue dots represent the predicted values for grid interpolations between $R_0=1.3$ to $R_0=2.05$ and $\theta_\mathrm{eq}=40^\circ$ to $\theta_\mathrm{eq}=95^\circ$.}
\label{fig:NNB-prediction}
\end{figure}

The parameter values obtained from $NN_{\tau}$ and $NN_{\beta}$ are visualized in Figures~\ref{fig:NNB-prediction}(a)~and~(b), respectively.
The parameter values obtained from PINNs are denoted by green and red circles, indicating the training data for $NN_{\beta}$ and $NN_{\tau}$, respectively. The parameter values generated by the networks are depicted as light orange and light blue dots. Visually, it is evident that these dots have filled the gaps between parameters that were absent in our LAMMPS dataset.
\section{Symbolic regression}\label{Symbolic_regression}

In this section, we use symbolic regression to find the explicit relation between the ODE parameters, and the initial radii and equilibrium contact angles.
Symbolic regression is a technique used in empirical modelling to discover mathematical expressions or symbolic formulas that best fit a given dataset~\cite{billard2002symbolic-incomplete}.
The process of symbolic regression involves searching a space of mathematical expressions to find the equation that best fits the data. The search is typically guided by a fitness function that measures the goodness of fit between the equation and the data. The fitness function is optimized using various techniques such as genetic algorithms, gradient descent, or other optimization algorithms. The equations discovered by symbolic regression are expressed in terms of familiar (i.e., more common) mathematical functions and variables, which can be easily understood and interpreted by humans. 

In this study, we used the Python library \texttt{gplearn} for symbolic regression~\cite{stephens2016genetic}.
As discussed in Section~\ref{grid_points}, in order to have more accurate formulation for the ODE parameters before using symbolic regression, we trained two networks, $NN_{\beta}$ and $NN_{\tau}$, using the discovered values $\beta(R_0,\theta_\mathrm{eq})$, and $\tau(R_0,\theta_\mathrm{eq})$ enabling us to predict the parameter values for grid interpolations where no corresponding data points were available.
The predicted parameters from both PINNs and the $NN_{\beta}$ and $NN_{\tau}$ networks  are fed through the symbolic regression model to discover a mathematical formulation for each parameter. For this purpose, $\theta_\mathrm{eq}$ and $R_0$ are fed as inputs, and $\eta$, $\beta$, and $\tau$ are the outputs. We set the population size to $5,000$ and evolve $20$ generations until the error is close to $1\%$. Since the equation consists of basic operations such as addition, subtraction, multiplication, and division, we do not require any custom functions. 

The following results of symbolic regression can be substituted in Equation~\eqref{ODE},
\begin{align}\label{parameters} 
    &\eta =  -0.255     \nonumber    \\
    &\beta = 0.283 + 0.27 \left(\frac{\theta_\mathrm{eq}}{d}\right)  \\
    &\tau =  6.13 \left(\frac{d}{\theta_\mathrm{eq}} + 1\right), \nonumber
\end{align}
where $d$ is the initial size of the droplet in mDPD units.

\begin{figure}[!htp]
\centering
\includegraphics[width=0.49\textwidth]{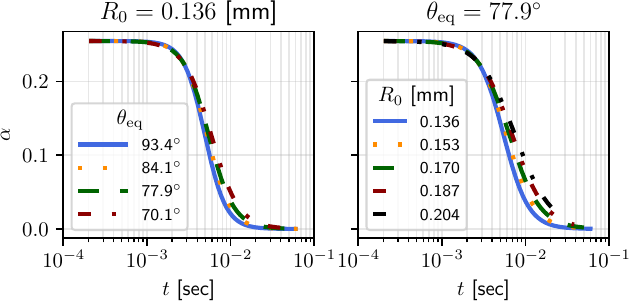}
\caption{The behaviour of $\alpha$ from RHS of Equation~\eqref{ODE} with parameters from Equation~\eqref{parameters}. Left: different contact angles with fixed initial radius $R_0=0.136$\,mm. Right: varying initial radii with fixed contact angle $\theta_\mathrm{eq}=77.9^\circ$.}
\label{discoverd-alpha}
\end{figure}

Figure~\ref{discoverd-alpha} shows the history of $\alpha(t, R_0, \theta_\mathrm{eq})$ using the the ODE (Equation~\eqref{ODE}) with parameters from Equation~\eqref{parameters}. The conversion between $d$ in mDPD units and $R_0$ in physical unit is $R_{0} = d\times1.701\times 10^{-2}$\,mm. 

\begin{figure}
\centering
\includegraphics[width=0.49\textwidth]{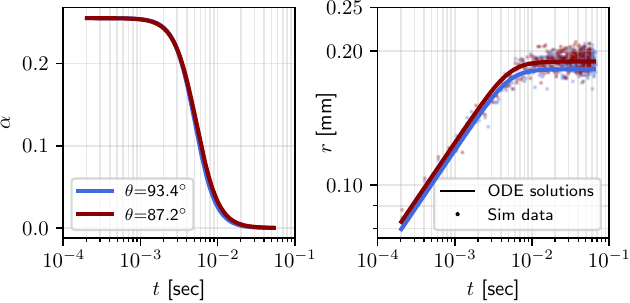}
\caption{The left figure illustrates the behavior of the parameter $\alpha$ using Equation~\eqref{ODE} for $R_{0} = 0.127$\,mm, which falls outside the range of the initial drop sizes used for training the networks.
On the right panel, the simulation data and the solution obtained from solving the ODE (Equation~\eqref{ODE}) with parameters from symbolic regression, Equation~\eqref{parameters} are shown.}
\label{discoverd-alpha-R0=127}
\end{figure}

In Figure~\ref{discoverd-alpha-R0=127}, the left panel depicts the values of $\alpha$ for an initial drop size of $R_0 = 0.127$\,mm and contact angles $\theta_\mathrm{eq}=93.4^\circ$ and $\theta_\mathrm{eq}=87.2^\circ$. The right panel compares the solution of Equation~\eqref{ODE} using the discovered parameters, Equation~\ref{parameters}, with the simulation data. The figure demonstrates the agreement between the ODE solution and the actual simulation results for this particular, unseen data set. It is important to note that this particular drop size lies outside the training interval for initial drop sizes $[0.136, 0.204]$\,mm. 

\section{Bayesian Physics-informed Neural Network: B-PINN results}\label{B-PINN}

Bayesian Physics-Informed Neural Networks (B-PINNs) integrate the traditional PINN framework with Bayesian Neural Networks (BNNs)~\cite{bykov2021explaining} to enable quantification of uncertainty in predictions~\cite{yang2021b}.
This framework combines the advantages of BNNs \cite{bishop1997bayesian} and PINNs to address both forward and inverse nonlinear problems. By choosing a prior over the ODE and network parameters, and by defining a likelihood function, one can find posterior distributions, using Bayes's theorem. 
B-PINNs offer a robust approach for handling problems containing uncorrelated noise, and they provide aleatoric and epistemic uncertainty quantification on the parameters of neural networks and ODEs.

The BNN component of the prior adopts Bayesian principles by assigning probability distributions to the weights and biases of the neural network. To account for noise in the data, we add noise to the likelihood function. By applying Bayes' rule, we can estimate the posterior distribution of the model and the ODE parameters. This estimation process enables the propagation of uncertainty from the observed data to the predictions made by the model.
We write Equation~\eqref{ODE} as
\begin{subequations}
\label{BODE}
\begin{align}
& \mathcal{N}_t(r; \bm{\lambda}) = f(t),\quad t \in \mathbb{R}^+ \\  
& \mathcal{I} (r, \bm{\lambda}) = r_0, \quad t=0,
\end{align}    
\end{subequations}
where $\bm{\lambda} = [\eta,~\beta, \tau]^\top$ is a vector of the parameters of the ODE (Equation~\eqref{ODE}), and  $\mathcal{N}_t$ is a general differential operator.  $f(t)$ is the forcing term, and $\mathcal{I}$ is the initial condition. This problem is an inverse problems, $\bm{\lambda}$  is inferred
from the data with estimates on aleatoric  and epistemic uncertainties.
The likelihoods of simulation data and ODE parameters are given as
\small
\begin{align}\label{liklihood}
 \nonumber
\!\! P(\mathcal{D} \! \mid \! \boldsymbol{\theta}, \bm{\lambda}) & \! = \! P\left(\mathcal{D}_{r} \mid \boldsymbol{\theta}\right) P\left(\mathcal{D}_f \mid \boldsymbol{\theta}, \bm{\lambda}\right) P\left(\mathcal{D}_{\mathcal{I}} \mid \boldsymbol{\theta}, \bm{\lambda}\right),~\text{where} \\  \nonumber
\! \! \! \! \! P\! \left(\mathcal{D}_{r} \! \mid \! \boldsymbol{\theta}, \bm{\lambda}\right) & \! = \!\prod_{i=1}^{N_{r}} \!  \frac{1}{\sqrt{2 \pi \sigma_r^{(i)^2}}} \exp \! \! \left[\! -\frac{\left({r}(\boldsymbol{t}_r^{(i)} ; \boldsymbol{\theta}, \boldsymbol{\lambda})-\bar{r}^{(i)}\right)^2}{2 \sigma_r^{(i)^2}}\right]\!\!, \\  \nonumber
\! \! \! \! \! P\! \left(\mathcal{D}_f \! \mid \! \boldsymbol{\theta}, \bm{\lambda}\right) & \! = \!\prod_{i=1}^{N_f} \!  \frac{1}{\sqrt{2 \pi \sigma_f^{(i)^2}}} \exp \! \! \left[\! -\frac{\left(f(\boldsymbol{t}_f^{(i)} ; \boldsymbol{\theta}, \boldsymbol{\lambda})-\bar{f}^{(i)}\right)^2}{2 \sigma_f^{(i)^2}}\right]\!\!, \\  
\!\! \! \! \! \! P\!\left(\mathcal{D}_\mathcal{I} \! \mid \! \boldsymbol{\theta}, \boldsymbol{\lambda}\right) & \! = \!\prod_{i=1}^{N_\mathcal{I}} \! \frac{1}{\sqrt{2 \pi \sigma_\mathcal{I}^{(i)^2}}} \exp \! \! \left[ \!-\frac{\left(\mathcal{I}(\boldsymbol{t}_i^{(i)} ; \boldsymbol{\theta}, \boldsymbol{\lambda}) \! - \! \bar{\mathcal{I}}^{(i)}\right)^2}{2 \sigma_{\mathcal{I}}^{(i)^2}}\right]\!\!,
\end{align}
\normalsize
where $D\! = \!D_r \cup D_f \cup D_{\mathcal{I}}$ with 
$\mathcal{D}_r\! = \!\left\{\left({\ln t}_r^{(i)}, \bar{\ln r}^{(i)}\right)\right\}_{i=1}^{N_r}$, 
$\mathcal{D}_f\! = \!\left\{\left(t_f^{(i)}, f^{(i)}\right)\right\}_{i=1}^{N_f}$, 
$\mathcal{D}_\mathcal{I}\! = \!\left\{\left(t_\mathcal{I}^{(i)}, \mathcal{}{I}^{(i)}\right)\right\}_{i=1}^{N_\mathcal{I}}$ are scattered noisy measurements.
The joint posterior of $[\bm{\theta}, \bm{\lambda}]$ is given as
\begin{align}\label{posterior}
\begin{aligned}
P(\boldsymbol{\theta}, \boldsymbol{\lambda} \mid \mathcal{D})&=\frac{P(\mathcal{D} \mid \boldsymbol{\theta}, \boldsymbol{\lambda}) P(\boldsymbol{\theta}, \boldsymbol{\lambda})}{P(\mathcal{D})}\\
&\simeq P(\mathcal{D} \mid \boldsymbol{\theta}, \boldsymbol{\lambda}) P(\boldsymbol{\theta}, \boldsymbol{\lambda})\\
&=P(\mathcal{D} \mid \boldsymbol{\theta}, \boldsymbol{\lambda}) P(\boldsymbol{\theta}) P(\boldsymbol{\lambda}).
\end{aligned}
\end{align}

To sample the parameters from the the posterior probability distribution defined by Equation~\eqref{posterior}, we utilized the Hamiltonian Monte Carlo (HMC) approach~\cite{radivojevic2020modified}, which is an efficient Markov Chain Monte Carlo (MCMC) method~\cite{brooks1998markov}.
For a detailed description of the method, 
please see, e.g., Refs.~\cite{neal2011mcmc, neal2012bayesian, graves2011practical}. To sample the posterior probability distribution, however, variational inference~\cite{blei2017variational} could be also used. In variational inference, the posterior density of the unknown parameter vector is approximated by another parameterized density function, which is restricted to a smaller family of distributions~\cite{yang2021b}. To compute the uncertainty in the ODE parameters by using B-PINN, a noise of $5 \%$ was added to the original data set. The noise was sampled from a normal distribution with a mean of $0$ and standard deviation of $\pm 1$. 

Here, the neural network model architecture comprises of two hidden layers, each containing 50 neurons. The network takes time ($t$) as the input, and generates a droplet radius $r(t)$ as the output. Additionally, we include a total of 2,000 burn-in samples.

The computational expense of B-PINNs compared to traditional neural networks primarily arises from the iterative nature of Bayesian inference and the need to sample from the posterior distribution. B-PINNs involve iterative Bayesian inference, where the posterior distribution is updated iteratively based on observed data. This iterative process requires multiple iterations to converge to a stable solution, leading to increased computational cost compared to non-iterative methods. Moreover, B-PINNs employ sampling-based algorithms such as MCMC or variational inference to estimate the posterior distribution of the model parameters. These algorithms generate multiple samples from the posterior distribution, which are used to approximate uncertainty and infer calibrated parameters. 

Sampling from the posterior distribution can be computationally expensive, particularly for high-dimensional parameter spaces or complex physics models. Furthermore, B-PINNs often require running multiple forward simulations of the physics-based model for different parameter samples. Each simulation represents a potential configuration of the model parameters. Since physics-based simulations can be computationally intensive, conducting multiple simulations significantly increases the computational cost of training B-PINNs. Achieving a high acceptance rate for posterior samples, especially for high-dimensional data, demands running a large number of simulations. This further adds to the computational complexity. 

\begin{figure}
\centering
\includegraphics[width=0.49\textwidth]{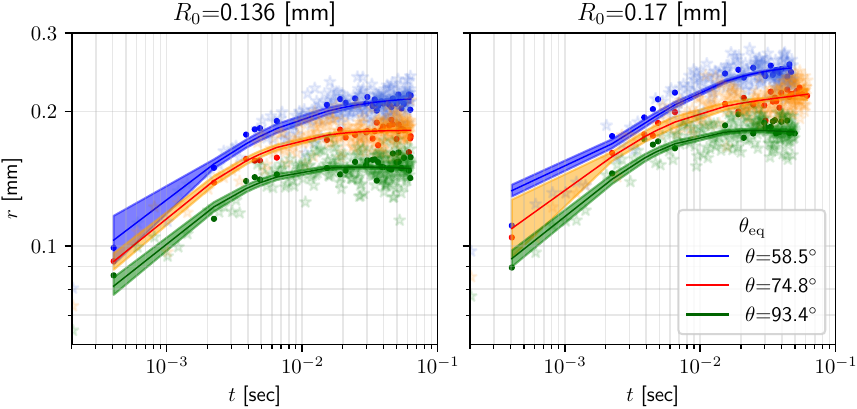}
\caption{The mean and uncertainty (mean $\pm$ 2\, standard deviation) of B-PINN predictions of the spreading radii history are given as solid lines and shaded regions, respectively.
The test simulation data is depicted by solid circles and training data  is indicated by stars.
This analysis is carried out for two different initial drop sizes, namely $R_0 = 0.137$\,mm and $0.170$\,mm, for three equilibrium contact angles.}
\label{fig:B_PINN-prediction}
\end{figure}

Due to the computational expense associated with B-PINNs, we opt to use the method selectively for a few cases only.
Utilizing the insights gained from PINNs prediction and symbolic regression, specifically the known value of $\eta = -0.255$, we can leverage the power of B-PINNs to uncover and ascertain the values of the parameters $\beta$ and $\tau$. Figure~\ref{fig:B_PINN-prediction} showcases the comparison between the mean values of the radii $\tilde{r}(t)$ predicted by B-PINNs represented by solid lines, the corresponding standard deviations denoted by highlighted regions, and the simulation data used for training is presented as stars, while the test data is indicated by colored circles. The horizontal axes represent time, while the vertical axes depict the spreading radii $\tilde{r}(t)$. This comparative analysis is conducted for two distinct initial drop sizes, namely $R_0 = 0.137$\,mm and $0.170$\,mm, considering various equilibrium contact angles. 

\begin{figure}
\centering
\includegraphics[width=0.49\textwidth]{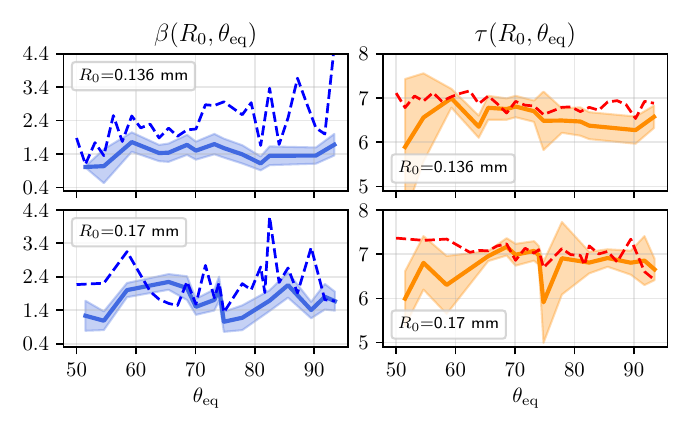}
\caption{Comparison between B-PINNs and PINNs discovered parameters for range of equilibrium contact angles and two initial radii. The mean values (solid lines) and the standard deviations (mean values $\pm$ 2\, standard deviations, shaded region) of $\beta$ (left panels) and $\tau$ (right panels). The dashed lines represent the parameters discovered by PINNs.}
\label{fig:B_PINN-parameters}
\end{figure}

Figure~\ref{fig:B_PINN-parameters} illustrates the mean values of the parameters $\beta$ and $\tau$ obtained using B-PINNs along with their corresponding standard deviations. The solid lines represent the average values of the discovered parameters, while the highlighted regions indicate the standard deviations. The parameters discovered by PINNs are represented by the dashed lines. On the left side, 
the vertical axes represent the values of $\beta$, while the
panels on the right side display the values of $\tau$. The results are presented for two initial drop sizes: $R_0=0.136$\,mm (top panels) and $R_0=0.17$\,mm (bottom panels). From the figure, it can be observed that the parameter $\beta$ exhibits a range of values between $1.0$ and $3.0$. On the other hand, the parameter $\tau$ fluctuates within the range of $5.0$ to $7.0$. These ranges provide insight into the variability and uncertainty associated with the estimated values of $\beta$ and $\tau$ obtained through the B-PINN methodology.

\begin{figure}
\centering
\includegraphics[width=0.49\textwidth]{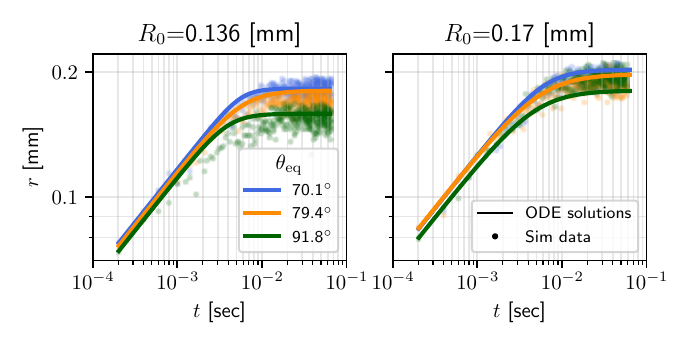}
\caption{Comparison between the ODE solution with parameters found by B-PINNs (solid lines), and the simulation radii (circles). Two initial drop sizes $R_0 = 0.137$\,mm and $0.170$\,mm and three equilibrium contact angles are shown.}
\label{fig:B_PINN-ode-solution}
\end{figure}

By comparing Figures~\ref{fig:abt0}~and~\ref{fig:B_PINN-parameters}, it becomes evident that the discovered parameters $\beta$ and $\tau$ using PINNs of B-PINNs frameworks
exhibit remarkable similarity. This striking similarity reinforces the efficacy and capability of our models in accurately identifying the parameters of the ODE described in Equation~\eqref{ODE}.
The close alignment between the discovered parameters in both figures demonstrates the robustness and reliability of our models. It highlights their ability to effectively capture the underlying dynamics and characteristics of the spreading behavior of CMAS, leading to accurate parameter estimation. This consistency and agreement between the PINN and B-PINN results provide further validation of the power and effectiveness of our modeling approaches in uncovering the true values of the parameters $\beta$ and $\tau$ in the ODE. 
Additionally, Equation~\eqref{ODE} with anticipated parameters obtained using B-PINNs is solved using \texttt{Odeint}. The results are presented in Figure~\ref{fig:B_PINN-ode-solution}, which provides a comparison between the simulated spreading radii $r(t)$ (circles) and the solution of the ODE (solid lines). This comparison is conducted for initial drop sizes of $R_0 = 0.136$\,mm and $0.17$\,mm considering different contact angles. 

\section{Conclusions}\label{Conclusion}

This study introduces a new approach to model the spreading dynamics of molten CMAS droplets. In the liquid state, CMAS is characterized by high viscosity, density, and surface tension. The main objective is to achieve a comprehensive understanding of the spreading dynamics by integrating the underlying physics into the neural network architecture. 

The study emphasizes the potential of PINNs in analyzing complex systems with intricate dynamics. 
To study the dynamics of CMAS droplets, we performed simulations using the mDPD method. By analyzing the simulation data and observing the droplet behavior,
we proposed a coarse parametric equation (Equation~\eqref{ODE}), which consists of three unknown parameters. This parametric equation aims to capture and describe the observed behavior of the CMAS droplets based on the simulation results.
Using the data from the mDPD simulations, the study employed the PINNs framework to determine the parameters of the equation. Symbolic regression was then utilized to establish the relationship between the identified parameter values, and the initial droplet radii and contact angles. As a result, a simplified ODE model was developed, accurately capturing the spreading dynamics. The model's parameters were explicitly determined based on the droplet's geometry and surface properties. Furthermore, B-PINNs were employed to assess the uncertainty associated with the model predictions, providing a comprehensive analysis of the spreading behavior of CMAS droplets.

Our findings extend beyond the specific case of CMAS droplets. The relationships uncovered and methods developed in this study have broader applications in understanding the spreading dynamics of droplets in general.
By leveraging the insights gained from this research, one can investigate and understand the behavior of droplets in diverse contexts, furthering our understanding of droplet spreading phenomena. This knowledge can potentially be used in developing strategies for effective droplet management and optimizing processes involving droplets in a wide range of practical applications.

\section*{Acknowledgments}
EK thanks the Mitacs Globalink Research Award Abroad and Western University's Science International Engagement Fund Award. MK thanks the Natural Sciences and Engineering Research Council of Canada (NSERC) and the Canada Research Chairs Program. The work was partially supported by DOE grant (DE-SC0023389). RBK would like to acknowledge the support received from the US Army Research Office Mathematical Sciences Division for this research through grant number W911NF-17-S-0002. LB and AG were supported by the US Army Research Laboratory 6.1 basic research
program in propulsion sciences. 
ZL and GK acknowledge the support from the AIM for Composites, an Energy Frontier Research Center funded by the U.S. Department of Energy (DOE), Office of Science, Basic Energy Sciences (BES) under Award \#DE-SC0023389. ZL also acknowledges support from the National Science Foundation (Grants OAC-2103967 and CDS\&E-2204011). Computing facilities were provided by the Digital Research Alliance of Canada (https://alliancecan.ca)
and the Center for Computation and Visualization, Brown University. The authors acknowledge the resources and support provided by Department of Defense Supercomputing Resource Center (DSRC) through use of "Narwhal" as part of the 2022 Frontier Project, Large-Scale Integrated Simulations of Transient Aerothermodynamics in Gas Turbine Engines. 


%

\end{document}